\documentclass[a4paper,10pt]{article}

\usepackage{amssymb}
\usepackage{amsmath}

\newcommand{\nl}{\newline}
\newcommand{\beq}{\begin{equation}}
\newcommand{\eeq}{\end{equation}}
\newcommand{\nec}{\newcommand}
\nec{\cts}{conformal transformations }
\nec{\fourg}{g_{\alpha\beta}} 
\nec{\grad}{\bigtriangledown}
\nec{\fourr}{^{(4)}R}
\nec{\detg}{^{(4)}g}
\nec{\eins}{\Biggl(R_{\alpha\beta} - \frac{1}{2}g_{\alpha\beta}R\Biggr)}
\nec{\kk}{K^{ab}K_{ab}}
\nec{\bec}{\begin{center}}
\nec{\eec}{\end{center}}
\nec{\bb}{B^{ab}B_{ab}}
\nec{\rp}{R - 8\frac{\grad^{2}\psi}{\psi}}
\nec{\pipi}{\pi^{ab}\pi_{ab}}
\nec{\beqq}{\begin{equation*}}
\nec{\eeqq}{\end{equation*}}
\nec{\V}{V(\psi)}
\nec{\delg}{\frac{\partial g_{ab}}{\partial t}}
\nec{\wt}{\widetilde}
\nec{\gt}{\longrightarrow}
\nec{\wh}{\widehat}
\begin{document}

\baselineskip=.65cm
\begin{center}
\noindent{\LARGE\bf{Scale-invariance in gravity and implications for the cosmological constant}}

\noindent{Bryan Kelleher}
\\
\noindent{\emph{Physics Department, University College, Cork, Ireland}}
\\
\noindent{E-mail: bk@physics.ucc.ie}
\end{center}
\renewcommand{\abstractname}{}
\begin{abstract}
\noindent{Recently a scale invariant theory was constructed by imposing a 
conformal symmetry on general relativity. The imposition of this symmetry 
changed the configuration space from \emph{superspace} - the space of all 
Riemannian $3$-metrics modulo diffeomorphisms - to \emph{conformal superspace} 
- the space of all Riemannian $3$-metrics modulo diffeomorphisms \emph{and} 
conformal transformations. However, despite numerous attractive features, the 
theory suffers from at least one major problem: the volume of the universe is 
no longer a dynamical variable. In attempting to resolve this problem a new 
theory is found which has several surprising and atractive features from both 
quantisation and cosmological perspectives. Furthermore, it is an extremely 
restrictive theory and thus may provide testable predictions quickly and 
easily. One particularly interesting feature of the theory is the resolution of
 the cosmological constant problem.}
\end{abstract}

\section{Introduction}
Despite many promising features the scale invariant gravity theory - conformal 
gravity - recently proposed in \cite{abfom} there is at least one major 
drawback. We can find the time derivative of the volume quite easily and get 
that it is proportional to $tr\pi$ and thus is zero. That is, the volume does 
not change and so the theory predicts a static universe and we cannot have 
expansion. This is quite a serious problem as the prediction of expansion in GR
 is considered to be one of the theory's greatest achievements. We are left 
with the following options:\nl
\\
(a) Abandon the theory; \nl
(b) Find a new explanation of the red-shift (among other things); \nl
(c) Amend the theory to recover expansion. \nl
\\
The first option seems quite drastic and the second, while certainly the most 
dramatic, also seems to be the most difficult. Thus, let's consider option (c).

\subsection{Resolving The problem(s)}
The notation used here will be the same as that used in \cite{bk}. In this 
notation the Lagrangian of the original theory is
\beq \label{lag1} \mathcal{L} = N\sqrt{g}\psi^{4}\biggl(\rp + B_{ab}B^{ab} - 
(trB)^{2}\biggr) \eeq
where $B_{ab}$ is the analogue of the extrinsic curvature. It is given by
\beq  B_{ab} = - \frac{1}{2N}\biggl(\delg - (KN)_{ab} - \theta 
g_{ab}\biggr) \eeq
The constraint
\beq tr\pi = 0 \eeq
arises from variation with respect to $\theta$ and so it is here that we shall 
make a change. Let's naively change the form of $B_{ab}$ to
\beq  \label{newb} B_{ab} = - \frac{1}{2N}\biggl(\delg - (KN)_{ab} - 
\grad_{c}\xi^{c}g_{ab}\biggr) \eeq
but we keep the original form for the Lagrangian equation (\ref{lag1}).
\\
Let's vary the action with respect to $\xi^{c}$. We get
\beq \begin{split} \delta\mathcal{L} = &
N\sqrt{g}\psi^{4}\biggl(2B^{ab} - 2trBg^{ab}\biggr)\delta B_{ab} \\ = & 
2N\sqrt{g}\psi^{4}\biggl(B^{ab} - 2trBg^{ab}\biggr)
\biggl(\frac{-1}{2N}\biggr)\biggl(-\grad_{c}\delta\xi^{c}g_{ab}\biggr) \\ = & 
-2\sqrt{g}\psi^{4}trB\grad_{c}\delta\xi^{c} \end{split} \eeq
Integrating by parts gives
\beq \delta\mathcal{L} = 2\sqrt{g}\grad_{c}(trB\psi^{4})\delta{\xi^{c}} \eeq
and so
\beq \grad_{c}(trB\psi^{4}) = 0 \eeq
(This will become the constant mean curvature (CMC) condition $\grad_{c}tr\pi =
 0$ later.)
\\
There is still one more equation which is found by varying with respect to 
$\psi$ - the so-called lapse fixing equation. However, since we have the same 
form for $\mathcal{L}$ as in the original theory our lapse-fixing equation is 
unchanged and as a result, the constraint is not propagated unless $tr\pi = 0$.
 Thus we haven't gained anything. We need a further change.
\\
It will prove instructive to split $B_{ab}$ into its trace and tracefree parts.
(The reason for this will become clear quite soon.) We label the tracefree part
 as $S_{ab}$. Thus we have
\beq B_{ab} = S_{ab} + \frac{1}{3}g_{ab}trB \eeq
We shall retain the new form of $B_{ab}$ as defined above in (\ref{newb}) all 
the same. The Lagrangian now reads
\beq \mathcal{L} = N\sqrt{g}\psi^{4}\biggl(\rp + S_{ab}S^{ab} - 
\frac{2}{3}(trB)^{2}\biggr) \eeq
We still need to make one further change. We'll simply stick in an additional 
$\psi$ term to the $trB$ part. (This is equivalent to redefining our conformal 
transformation so that $S_{ab}$ and $trB$ transform in different ways.) The 
Lagrangian takes the form
\beq \mathcal{L} = N\sqrt{g}\psi^{4}\biggl(\rp + S_{ab}S^{ab} - 
\frac{2}{3}\psi^{n}(trB)^{2}\biggr) \eeq
\\
Before we continue, one interesting point about $S_{ab}$ is the following. We 
have
\beq S_{ab} = B_{ab} - \frac{1}{3}g_{ab}trB \eeq
Let's write this out explicitly. We have
\beq S_{ab} = - \frac{1}{2N}\biggl(\delg - (KN)_{ab} - 
\grad_{c}\xi^{c}g_{ab}\biggr) + 
\frac{1}{3}\frac{1}{2N}g_{ab}\biggl(g^{cd}\frac{\partial 
g_{cd}}{\partial t} - g^{cd}(KN)_{cd} - 3\grad_{c}\xi^{c}\biggr) \eeq
Splitting this up further gives
\beq S_{ab} = - \frac{1}{2N}\biggl(\delg - (KN)_{ab}\biggr) + 
\frac{1}{2N}g_{ab}\grad_{c}\xi^{c} - 
\frac{1}{3}g_{ab}\biggl(g^{cd}\frac{\partial g_{cd}}{\partial t} - 
g^{cd}(KN)_{cd}\biggr) -\frac{1}{2N}g_{ab}\grad_{c}\xi^{c} \eeq
and with a simple cancellation
\beq S_{ab} = - \frac{1}{2N}\biggl(\delg - (KN)_{ab}\biggr) - 
\frac{1}{3}g_{ab}\biggl(g^{cd}\frac{\partial g_{cd}}{\partial t} - 
g^{cd}(KN)_{cd}\biggr) \eeq
Of course, this is
\beq S_{ab} = K_{ab} - \frac{1}{3}g_{ab}trK \eeq
That is, $S_{ab}$ is the tracefree part of the extrinsic curvature and is 
\emph{independent} of \emph{any} conformal fields.
\\ \\
Let us find $\pi^{ab}$. This is done as usual by varying with respect to 
$\delg$. We get
\beq \begin{split} \delta\mathcal{L} & =  
2N\sqrt{g}\psi^{4}\biggl(2S^{ab}\delta S_{ab} - 
\frac{4}{3}\psi^{n}trBg^{ab}\delta B_{ab}\biggr) 
\\& = 2N\sqrt{g}\psi^{4}\biggl(S^{ab}\biggl(\delta B_{ab} - 
\frac{1}{3}g_{ab}g^{cd}\delta B_{cd}\biggr) - 
\frac{2}{3}\psi^{n}trBg^{ab}\delta B_{ab}\biggr)
\\& = 2N\sqrt{g}\psi^{4}\biggl(S^{ab} - 
\frac{2}{3}\psi^{n}trBg^{ab}\biggr)\delta B_{ab}
\\& = - \sqrt{g}\psi^{4}\biggl(S^{ab} - \frac{2}{3}\psi^{n}S^{ab}trB\biggr)
\delta \delg \end{split} \eeq
Thus,
\beq \pi^{ab} = -\sqrt{g}\psi^{4}S^{ab} + 
\frac{2}{3}\sqrt{g}\psi^{n+4}g^{ab}trB \eeq
Splitting $\pi^{ab}$ into its trace and tracefree parts will further clear 
things up. We'll label the split as
\beq \pi^{ab} = \sigma^{ab} + \frac{1}{3}g^{ab}tr\pi \eeq
Thus the tracefree part of $\pi^{ab}$ is
\beq \sigma^{ab} = -\sqrt{g}\psi^{4}S^{ab} \eeq
and the trace is given by
\beq tr\pi = 2\psi^{n+4}trB \eeq
Note that our value of $n$ is undefined as yet.
\\ \\
The constraints are found by varying with respect to $\xi^{c}$, $\psi$, $N$ and
$N^{a}$. The conformal constraint and the lapse-fixing equation are given by 
varying with respect to $\xi^{c}$ and $\psi$ respectively. These give
\beq \grad_{c}tr\pi = 0 \eeq
and
\beq N\psi^{3}\biggl(R - 7\frac{\grad^{2}\psi}{\psi}\biggr) - 
\grad^{2}(N\psi^{3}) + \frac{(trp)^{2}\psi^{7}}{4} = 0 \eeq
respectively. From the variation with respect to $N$ we get
\beq S_{ab}S^{ab} - \frac{2}{3}\psi^{n}(trB)^{2} - g\psi^{8}\biggl(\rp\biggr) =
 0 \eeq
which in terms of the momentum is
\beq \label{hami} \sigma_{ab}\sigma^{ab} - \frac{1}{6}\psi^{-n}(tr\pi)^{2} - 
g\psi^{8}\biggl(\rp\biggr) = 0 \eeq
and finally, from the variation with respect to $N^{a}$ we get
\beq \label{xxx} \grad_{b}\pi^{ab} = 0 \eeq
We require conformal invariance in our constraints. Under what conditions is 
the momentum constraint (\ref{xxx}) invariant? The tracefree part of the 
momentum, $\sigma^{ab}$, has a natural weight of $-4$ (from the original 
theory). That is
\beq \sigma^{ab} \gt \omega^{-4}\sigma^{ab} \eeq
If $tr\pi = 0$ then we have conformal invariance. If not however, we require 
various further conditions. We need
\beq \grad_{b}\sigma^{ab} = 0 \eeq
\beq \grad_{c}tr\pi = 0 \eeq
and that
\beq trp = \frac{tr\pi}{\sqrt{g}} \gt trp \eeq
under a conformal transformation. In our theory we have the first two 
conditions emerging directly and naturally from the variation. Thus we simply 
define $trp$ to transform as a conformal scalar as required. With this done our
 momentum constraint is conformally invariant.
\\ \\
Transforming the constraint (\ref{hami}) gives
\beq \sigma_{ab}\sigma^{ab} - \frac{1}{6}\psi^{-n}g(trp)^{2}\omega^{12 + n} - 
g\psi^{8}\biggl(\rp\biggr) = 0 \eeq
and so we must have $n = -12$ for conformal invariance.
The constraint then becomes
\beq \sigma_{ab}\sigma^{ab} - \frac{1}{6}\psi^{12}g(trp)^{2} - 
g\psi^{8}\biggl(\rp\biggr) = 0 \eeq
This is \emph{exactly} the Lichnerowicz-York equation from GR \cite{yor}. 
However, we have found it directly from a variational procedure.
\\ \\
Thus we have determined the unique value of $n$ and our constraints are
\beq \label{c1new} \sigma_{ab}\sigma^{ab} - \frac{1}{6}\psi^{12}(tr\pi)^{2} - 
g\psi^{8}\biggl(\rp\biggr) = 0 \eeq
\beq \label{c2new} \grad_{b}\pi^{ab} = 0 \eeq
\beq \label{c3new}\grad_{c}tr\pi = 0 \eeq
Thus we have found the exact constraints of the York method \cite{yor} directly from a 
variational procedure. This is quite novel. We also have a lapse fixing equation
\beq \label{c4new} N\psi^{3}\biggl(R - 7\frac{\grad^{2}\psi}{\psi}\biggr) - 
\grad^{2}(N\psi^{3}) + \frac{(trp)^{2}\psi^{7}}{4} = 0 \eeq
It turns out (as we shall show later) that this condition enforces propagation of the 
constraint (\ref{c3new}).  
\\ \\
Let's proceed to the Hamiltonian formulation.
\section{The Hamiltonian Formulation}
The earlier expression for $\pi^{ab}$ can be inverted to get $\delg$. We get
\beq \label{gstuff} \delg = \frac{2N}{\sqrt{g}\psi^{4}}\biggl(\sigma_{ab} - 
\frac{1}{6}g_{ab}tr\pi\psi^{12}\biggr) + (KN)_{ab} + 
g_{ab}\grad_{c}\xi^{c} \eeq
The Hamiltonian may then be found in the usual way. We get
\beq \mathcal{H} = 
\frac{N}{\sqrt{g}\psi^{4}}\biggl[\sigma^{ab}\sigma_{ab} - 
\frac{1}{6}(tr\pi)^{2}\psi^{12} - g\psi^{8}\biggl(R - 
8\frac{\grad^{2}\psi}{\psi}\biggr)\biggr] -
 2N_{a}\grad_{b}\pi^{ab} - \xi^{c}\grad_{c}tr\pi \eeq
As a consistency check let's find $\delg$ from this by varying with respect to 
$\pi^{ab}$. We get exactly equation (\ref{gstuff}) again. Thus, all is well. We may do all the usual variations here to get 
the constraints. Varying the Hamiltonian with respect to $g_{ab}$ gives us our 
evolution equation for $\pi^{ab}$. We get
\begin{equation}
\begin{split}
\frac{\partial\pi^{ab}}{\partial t} = & - 
N\sqrt{g}\psi^{4}\biggl(R^{ab} - g^{ab}\biggl(\rp\biggr)\biggr) \\ & - 
\frac{2N}{\sqrt{g}\psi^{4}}\biggl(\pi^{ac}\pi^{b}_{\;\;c} - \frac{1}{3}\pi^{ab}
tr\pi - \frac{1}{6}\pi^{ab}tr\pi\psi^{12}\biggr) \\ & + \sqrt{g}\psi
\biggl(\grad^{a}\grad^{b}(N\psi^{3}) - g^{ab}\grad^{2}(N\psi^{3})\biggr) \\ & 
+ N\sqrt{g}\psi^{3}\biggl(\grad^{a}\grad^{b}\psi + 3g^{ab}\grad^{2}\psi\biggr) 
 \\ & + 4g^{ab}\sqrt{g}\grad_{c}(N\psi^{3})\grad^{c}\psi - 
6\sqrt{g}\grad^{(a}(N\psi^{3})\grad^{b)}\psi
\\ & + \grad_{c}\biggl(\pi^{ab}N^{c}\biggr) - \pi^{bc}\grad_{c}N^{a} - 
\pi^{ac}\grad_{c}N^{b} \\ & - \biggl(\pi^{ab} - 
\frac{1}{2}g^{ab}tr\pi\biggr)\grad_{c}\xi^{c} 
\end{split}
 \eeq
We may use the evolution equations to find $\frac{\partial trp}{\partial t}$ 
quite easily. (Of course, we need the evolution equations to propagate all of 
the constraints. We will deal with the others later.) We find that
\beq \frac{\partial trp}{\partial t} = 0 \eeq
using the lapse-fixing equation. Thus we have that $trp =$ constant both 
spatially \emph{and} temporally!! (It is constant spatially since the 
densitised momantum $\pi^{ab}$ is covariantly constant.
\beq 0 = \grad_{c}tr\pi = \frac{1}{\sqrt{g}}\grad_{c}trp \eeq
Thus
\beq trp_{,c} = 0. \eeq)
We could proceed to check propagation of the constraints here but it will be 
easier and more instructive to do a little more work first.
\\ \\
Since $trp$ is identically a constant our dynamical data consists of $g_{ab}$ 
and $\sigma^{ab}$. Thus, it may prove useful to have an evolution equation for 
$\sigma^{ab}$ rather than the full $\pi^{ab}$. It is reasonably straightforward
 to do this. Firstly we note that
\beq \frac{\partial\sigma^{ab}}{\partial t} = 
\frac{\partial\pi^{ab}}{\partial t} - 
\frac{1}{3}\frac{\partial g^{ab}tr\pi}{\partial t} \eeq 
Working through the details gives us
\beq \begin{split} \frac{\partial\sigma^{ab}}{\partial t} = & - 
N\sqrt{g}\psi^{4}\biggl(R^{ab} - \frac{1}{3}g^{ab}\biggl(\rp\biggr)\biggr) - 
\frac{2N}{\sqrt{g}\psi^{4}}\sigma^{ac}\sigma^{b}_{\;\;c} \\ & + \sqrt{g}\psi
\biggl(\grad^{a}\grad^{b}(N\psi^{3}) - 
\frac{1}{3}g^{ab}\grad^{2}(N\psi^{3})\biggr) \\ & 
+ N\sqrt{g}\psi^{3}\biggl(\grad^{a}\grad^{b}\psi + 
\frac{7}{3}g^{ab}\grad^{2}\psi\biggr) 
 \\ & + 4g^{ab}\sqrt{g}\grad_{c}(N\psi^{3})\grad^{c}\psi - 
6\sqrt{g}\grad^{(a}(N\psi^{3})\grad^{b)}\psi 
\\ & + \grad_{c}\bigl(\sigma^{ab}N^{c}\biggr) -  \sigma^{bc}\grad_{c}N^{a} - 
 \sigma^{ac}\grad_{c}N^{b} \\ & - \sigma^{ab}\grad_{c}\xi^{c} + 
\frac{N\psi^{8}}{3\sqrt{g}}\sigma^{ab}tr\pi \end{split} \eeq
\section{Jacobi Action}
In 1962 Baierlein, Sharp and Wheeler (BSW) \cite{bsw} constructed a 
Jacobi action for G.R. It was of the form
\begin{equation}
I = \int d\lambda \int \sqrt{g} \sqrt{R} \sqrt{T} d^3x,
\end{equation}
where the `kinetic energy' $T$ is 
\begin{eqnarray}
T &=& (g^{ac}g^{bd} - g^{ab}g^{cd})\nonumber\\& 
&\left({\partial g_{ab} \over
\partial
\lambda} - (KW)_{ab}\right)\left({\partial g_{cd} \over 
\partial\lambda} - (KW)_{cd}\right).
\end{eqnarray}
This action reproduces the standard Einstein equations in the 
Arnowitt-Deser-Misner form \cite{adm} with lapse $N = \sqrt{T/4R}$. We can also find the 
Jacobi action for the new theory. Recall the (3+1) Lagrangian,
\beq \mathcal{L} = N\sqrt{g}\psi^{4}\biggl(\rp + S^{ab}S_{ab} - 
\frac{2}{3}\psi^{-12}(trB)^{2}\biggr) \eeq
We can write this as
\beq \mathcal{L} = \sqrt{g}\psi^{4}\biggl[N\biggl(\rp\biggr) + 
\frac{1}{4N}\biggl(\Sigma^{ab}\Sigma_{ab} - 
\frac{2}{3}\psi^{-12}(tr\beta)^{2}\biggr)\biggr] \eeq
where $ \Sigma_{ab} = - 2NS_{ab}$ and $\beta_{ab} = -2NB_{ab}$. We now 
extremise with respect to $N$. This gives us,
\beq N = \underline{+}\frac{1}{2}\biggl(\Sigma^{ab}\Sigma_{ab} - 
\frac{2}{3}\psi^{-12}(tr\beta)^{2}\biggr)^{\frac{1}{2}}
\biggl(\rp\biggr)^{-\frac{1}{2}} \eeq
Substituting this back into the action gives us
\beq \label{nct} S = \underline{+}\int 
d\lambda\int\sqrt{g}\psi^{4}\sqrt{\rp}\sqrt{T}d^{3}x \eeq
where $ T = \biggl(\Sigma^{ab}\Sigma_{ab} - 
\frac{2}{3}\psi^{-12}(tr\beta)^{2}\biggr) $.
\\ \\
We can do all the usual variations here: $N^{a}$, $\xi^{c}$ and $\psi$. 
These give the momentum constraint, the conformal constraint and the 
lapse-fixing condition respectively. When we find the canonical momentum 
$\pi^{ab}$ we can ``square'' it to give the ``Hamiltonian constraint.''
\\ \\
So far, so good. We shall rarely use the Jacobi form of the action here but 
from a thin-sandwich point of view it is important and may well be of use in 
future work. Let's move on.
\section{Conformally Related Solutions}
In conformal superspace conformally related metrics are equivalent. 
Thus, conformally related solutions of this theory must be physically 
equivalent and so it is crucial that we have a natural way to relate 
such solutions. Suppose we start with initial data 
$\{g_{ab},\sigma^{ab},trp\}$ obeying the initial data conditions 
(\ref{c2new}) and (\ref{c3new}). We then solve (\ref{c1new}) for 
$\psi$.
\\ \\
Suppose instead that we start with the conformally related initial 
data $\{h_{ab},\rho^{ab},trp\} = 
\{\alpha^{4}g_{ab},\alpha^{-4}\sigma^{ab},trp\}$. 
These automatically satisfy the initial data conditions by the conformal 
invariance. We now solve the Hamiltonian constraint for the conformal ``field''
 $\chi$, say. Exactly as in \cite{bk} it can be shown that $\chi = 
\frac{\psi}{\alpha}$. Thus,
\beq \psi^{4}g_{ab} = \chi^{4}h_{ab} \eeq
and
\beq \psi^{-4}\sigma^{ab} = \chi^{-4}\rho^{ab} \eeq
Let us label these as $\wt{g_{ab}}$ and $\wt{\rho^{ab}}$ and put a hat over 
$trp$ also (for clarity). Thus a tilde over a quantity denotes the 
physical value of this quantity. It is very 
remarkable that we find not only a physical momentum, which is 
precisely analogous to the physical gauge-invariant electric field in 
Maxwellian theory, but also to a physical $g_{ab}$. This has no 
analogue in Maxwell and Yang--Mills, in which the vector-potential 
velocity $\dot{A}_k$ is gauge-corrected by the scalar potential, 
yielding the gauge-invariant electric field $\textbf E$, but $A_k$ 
itself retains irremovable gauge degeneracy.

The constraints become
\beq \wt{\sigma_{ab}}\wt{\sigma^{ab}} - \frac{1}{6}\wt{(tr\pi)}^{2} - 
\wt{g}\wt{R} = 0 \eeq
\beq \wt{\grad_{b}}\wt{\pi^{ab}} = 0 \eeq
\beq \wt{\grad_{c}}\wt{tr\pi} = 0 \eeq
\beq \wt{N}\wt{R} - \wt{\grad^{2}}\wt{N} + \frac{(\wt{trp})^{2}}{4} = 0 \eeq
\\
Consider GR in the CMC gauge. The constraints are
\beq \sigma_{ab}\sigma^{ab} - \frac{1}{6}(tr\pi)^{2} - gR = 0 \eeq
\beq \grad_{b}\pi^{ab} = 0 \eeq
\beq \grad_{c}tr\pi = 0 \eeq
Evolution of the CMC condition gives
\beq NR - \grad^{2}N + \frac{(trp)^{2}}{4} = C \eeq
The similarities are \emph{quite} striking.
\subsection{What of $\xi^{c}$?}
Precious little has been revealed about what $\xi^{c}$ may be or even how it 
transforms. This needs to be addressed. First let's recall that we demanded 
that
\beq trB \gt \omega^{-8}trB \eeq
under a conformal transformation. This will be enough to reveal the 
transformation properties of $\xi^{c}$. Taking the trace gives us
\beq trB = - \frac{1}{2N}\biggl(g^{ab}\delg - g^{ab}(KN)_{ab} - 
3\grad_{c}\xi^{c}\biggr) \eeq
Under a conformal transformation we get
\beq 
\begin{split}  \omega^{-8}trB = & - \frac{1}{2\omega^{2}N}\biggl(g^{ab}\delg + 
12\frac{\dot{\omega}}{\omega} - \omega^{-4}g^{ab}\biggl(\omega^{4}(KN)_{ab} + 
4\omega^{3}\omega_{,c}N^{c}g_{ab}\biggr) - 3\bar{\grad_{c}}\bar{\xi^{c}}\biggr)
\\ = & - \frac{1}{2\omega^{2}N}\biggl(g^{ab}\delg - g^{ab}(KN)_{ab} - 
3\grad_{c}\xi^{c}\biggr) \\ & - \frac{3}{2\omega^{2}N}\grad_{c}\xi^{c} + 
\frac{3}{2\omega^{2}N}\bar{\grad_{c}}\bar{\xi^{c}} - 
\frac{6}{\omega^{3}N}\biggl(\dot{\omega} - \omega_{,c}N^{c}\biggr) \\
= & \omega^{-2}trB + \frac{3}{2\omega^{2}N}\biggl(\bar{\grad_{c}}\bar{\xi^{c}}
 - 3\grad_{c}\xi^{c} - 
\frac{4}{\omega}\biggl(\dot{\omega} - \omega_{,c}N^{c}\biggr)\biggr) 
\end{split} \eeq
Thus,
\beq \frac{3}{2\omega^{2}N}\biggl(\bar{\grad_{c}}\bar{\xi^{c}}
 - 3\grad_{c}\xi^{c} - \frac{4}{\omega}
\biggl(\dot{\omega} - \omega_{,c}N^{c}\biggr)\biggr)
= - \frac{1}{\omega^{2}N}trB\biggl(1 - \omega^{-6}\biggr) \eeq
and so
\beq \label{txic} \bar{\grad_{c}}\bar{\xi^{c}} = \grad_{c}\xi^{c} + 
\frac{4}{\omega}\biggl(\dot{\omega} - \omega_{,c}N^{c}\biggr) - 
\frac{2N}{3}trB\biggl(1 - \omega^{-6}\biggr) \eeq
This tells us how things transform but not what $\xi^{c}$ itself actually is. 
We \emph{can} find this though.
\\ \\
Let's write the evolution equations in the physical representation. It can be 
verified that they are
\beq \frac{\partial \wt{g_{ab}}}{\partial t} = \frac{2\wt{N}}{\sqrt{\wt{g}}}
\biggl(\wt{\sigma_{ab}} - \frac{1}{6}\wt{g_{ab}}\wt{tr\pi}\biggr) + 
\wt{(KN)_{ab}} + \wt{g_{ab}}\wt{\grad_{c}}\wt{\xi^{c}} \eeq
and
\beq
\begin{split} \frac{\partial\wt{\sigma^{ab}}}{\partial t} = & - 
\wt{N}\sqrt{\wt{g}}\biggl(\wt{R^{ab}} - \frac{1}{3}\wt{g^{ab}}\wt{R}\biggr) - 
\frac{2\wt{N}}{\sqrt{\wt{g}}}
\wt{\sigma^{ac}}\wt{\sigma^{b}_{\;\;c}} \\ & + \sqrt{\wt{g}}
\biggl(\wt{\grad^{a}}\wt{\grad^{b}}\wt{N} - 
\frac{1}{3}\wt{g^{ab}}\wt{\grad^{2}}\wt{N}
\biggr) \\ & + \wt{\grad_{c}}\biggl(\wt{\sigma^{ab}}\wt{N^{c}}\biggr) - 
\wt{\sigma^{bc}}\wt{\grad_{c}}\wt{N^{a}} - 
\wt{\sigma^{ac}}\wt{\grad_{c}}\wt{N^{b}} 
\\ & + \frac{\wt{N}}{3\sqrt{\wt{g}}}\wt{\sigma^{ab}}\wt{tr\pi} - 
\wt{\sigma^{ab}}\wt{\grad_{c}}\wt{\xi^{c}}\end{split} \eeq
We require the evolution equations to propagate the constraints. However, when
 we check this it turns out that we are \emph{forced} to set 
$\wt{\grad_{c}}\wt{\xi^{c}}$ to zero. However, this means that we have
\beq \grad_{c}\xi^{c} + 
\frac{4}{\omega}(\dot{\psi} - \psi_{,c}N^{c}) - \frac{2N}{3}trB(1 - \psi^{-6}) 
= 0 \eeq
by (\ref{txic}). Thus we have
\beq \grad_{c}\xi^{c} =  - \frac{4}{\psi}(\dot{\psi} - \psi_{,c}N^{c}) + 
\frac{2N}{3}trB(1 - \psi^{-6}) \eeq
That is,
\beq \grad_{c}\xi^{c} = \theta + \frac{2N}{3}trB(1 - \psi^{-6}) \eeq
where $\theta$ is as in the original theory. Thus, the exact form of $\xi^{c}$ 
is determined. We needed $\grad_{c}\xi^{c}$ to be zero in the physical 
representation for constraint propagation and so we should check that this is 
the case with our newly found expression for $\grad_{c}\xi^{c}$. We can check 
this easily. In the physical representation $\theta = 0$ and $\psi = 1$. Thus, 
we do have that $\wt{\grad_{c}}\wt{\xi^{c}}$ is zero.
\\ \\
It is vital to note that this is strictly a \emph{POST-VARIATION} 
identification. If we use this form for $\xi^{c}$ in the action we will run 
into problems, not least an infinite sequence in the variation of $trB$ with 
respect to $\xi^{c}$. (This is because we would have $trB$ defined in terms of 
$trB$ itself.) We see that $\xi^{c}$ is intimately related with how $\psi$ 
changes from slice to slice.
\\ \\
Our constraints in the physical representation are
\beq \sigma_{ab}\sigma^{ab} - \frac{1}{6}(tr\pi)^{2} - gR = 0 \eeq
\beq \grad_{b}\pi^{ab} = 0 \eeq
\beq \grad_{c}tr\pi = 0 \eeq
\beq NR - \grad^{2}N + \frac{N(trp)^{2}}{4} = 0 \eeq
and our evolution equations are
\beq \label{blugh} \frac{\partial g_{ab}}{\partial t} = \frac{2N}{\sqrt{g}}
\biggl(\sigma_{ab} - \frac{1}{6}g_{ab}tr\pi\biggr) + (KN)_{ab} \eeq
and
\beq \label{yuch} \begin{split} \frac{\partial\sigma^{ab}}{\partial t} = & - 
N\sqrt{g}\biggl(R^{ab} - \frac{1}{3}g^{ab}R\biggr) - 
\frac{2N}{\sqrt{g}}\sigma^{ac}\sigma^{b}_{\;\;c} \\ & + \sqrt{g}
\biggl(\grad^{a}\grad^{b}N - \frac{1}{3}g^{ab}\grad^{2}N
\biggr) \\ & + \grad_{c}(\sigma^{ab}N^{c}) - 
\sigma^{bc}\grad_{c}N^{a} - 
\sigma^{ac}\grad_{c}N^{b} 
\\ & + \frac{N}{3\sqrt{g}}\sigma^{ab}tr\pi \end{split} \eeq
(The hats are removed for simplicity.) These are \emph{identical} to those in 
GR in the CMC gauge (with $trp$ a temporal constant).
\section{Topological Considerations}
In the original theory it was found that if the manifold is compact without 
boundary we get frozen dynamics. In this problematic case we can resolve the 
issue in much the same manner as with the original theory although, it is a 
\emph{little} more complicated this time.
\subsection{Integral Inconsistencies}
The root of the integral inconsistency is in the lapse-fixing equation. If we 
integrate this equation we find that the only solution is $N \equiv 0$. That 
is, we have frozen dynamics. The resolution to this in the original conformal 
theory was to introduce a volume term in the denominator of the Lagrangian. 
Actually, the key is to keep the Lagrangian homogeneous in $\psi$ using 
different powers of the volume. The volume of a hypersurface here is given by
\beq V = \int \sqrt{g}\psi^{6}\;d^{3}x \eeq
In the original theory the Lagrangian has an overall factor of $\psi^{4}$ and 
so we need to divide by $V^{2/3}$ to keep homogeneity in $\psi$. There is no 
such overall factor in the new theory and so it is not as straightforward. The 
key is to treat the two parts of the Lagrangian separately. We try
\beq \mathcal{L}_{1} = \frac{N\sqrt{g}\psi^{4}}{V^{n}}\biggl(R - 
8\frac{\grad^{2}\psi}{\psi} + S^{ab}S_{ab}\biggr) \eeq
and
\beq \mathcal{L}_{2} = -\frac{2}{3}\frac{N\sqrt{g}\psi^{-8}}{V^{m}}(trB)^{2} 
\eeq
and we determine $n$ and $m$ from the homogeneity requirement. Thus we have 
that $n = \frac{2}{3}$ and $m = -\frac{4}{3}$. Using this result our Lagrangian
is now
\beq \label{lag3} \mathcal{L} = \frac{N\sqrt{g}\psi^{4}}{V^{2/3}}\biggl(R - 
8\frac{\grad^{2}\psi}{\psi} + S^{ab}S_{ab} - 
\frac{2}{3}\psi^{-12}(trB)^{2}V^{2}\biggr) \eeq
\subsection{New Constraints}
We go about things in exactly the same manner as before. The momentum is found 
to be
\beq \label{rightok} \pi^{ab} = - \frac{\sqrt{g}\psi^{4}}{V^{2/3}}S^{ab} + 
\frac{2}{3}\sqrt{g}\psi^{-8}V^{4/3}g^{ab}trB \eeq
The constraints are (almost) unchanged. They are
\beq \sigma_{ab}\sigma^{ab} - \frac{\psi^{12}(tr\pi)^{2}}{6V^{2}} - 
\frac{g\psi^{8}}{V^{4/3}}\biggl(\rp\biggr) = 0 \eeq
\beq \grad_{b}\pi^{ab} = 0 \eeq
\beq \grad_{c}tr\pi = 0 \eeq
The lapse-fixing equation is
\beq \frac{N\sqrt{g}\psi^{3}}{V^{2/3}}\biggl(R - 7\frac{\grad^{2}\psi}{\psi}
\biggr) - \sqrt{g}\frac{\grad^{2}(N\psi^{3})}{V^{2/3}} - 
\sqrt{g}\frac{C\psi^{5}}{2V^{2/3}} + 
\sqrt{g}N\psi^{-9}(trB)^{2}V^{4/3} - 
\frac{2}{3}\sqrt{g}D\psi^{5}V^{4/3} = 0 \eeq
where
\beq C = \int \frac{N\sqrt{g}\psi^{4}}{V}\biggl(R - 
8\frac{\grad^{2}\psi}{\psi} + S^{ab}S_{ab}\biggr) \; d^{3}x \eeq
and
\beq D = \int \frac{N\sqrt{g}\psi^{-8}}{V}(trB)^{2}\;d^{3}x \eeq
The $C$ and $D$ terms result from the variations of the volume. Rearranging the
 lapse-fixing equation we get
\beq \frac{N\sqrt{g}\psi^{3}}{V^{2/3}}\biggl(R - 7\frac{\grad^{2}\psi}{\psi}
\biggr) - \sqrt{g}\frac{\grad^{2}(N\psi^{3})}{V^{2/3}} 
\sqrt{g}N\psi^{-9}(trB)^{2}V^{4/3} = 
\frac{\sqrt{g}\psi^{5}}{2V^{2/3}}\biggl(C + \frac{4}{3}DV^{2}\biggr) \eeq
Integrating across this expression gives \emph{no} problem. The inconsistency 
has been removed.
\section{The Hamiltonian Formulation}
We should consider the evolution equations again now that we have changed the 
action. First of all the momentum is now given by (\ref{rightok}). The new Hamiltonian is
\beq \mathcal{H} = 
\frac{NV^{2/3}}{\sqrt{g}\psi^{4}}\biggl[\sigma^{ab}\sigma_{ab} - 
\frac{(tr\pi)^{2}\psi^{12}}{6V^{2}} - \frac{g\psi^{8}}{V^{4/3}}\biggl(R - 
8\frac{\grad^{2}\psi}{\psi}\biggr)\biggr] -
 2N_{a}\grad_{b}\pi^{ab} - \xi^{c}\grad_{c}tr\pi \eeq
The evolution equations are then
\beq \delg = \frac{2NV^{2/3}}{\sqrt{g}\psi^{4}}\biggl(\sigma_{ab} - 
\frac{g_{ab}tr\pi\psi^{12}}{6V^{2}}\biggr) + (KN)_{ab} + 
g_{ab}\grad_{c}\xi^{c} \eeq
and
\beq
\begin{split} \frac{\partial\pi^{ab}}{\partial t} = & - 
\frac{N\sqrt{g}\psi^{4}}{V^{2/3}}\biggl(R^{ab} - 
g^{ab}\biggl(\rp\biggr)\biggr) \\ & - 
\frac{2NV^{2/3}}{\sqrt{g}\psi^{4}}\biggl(\pi^{ac}\pi^{b}_{\;\;c} - 
\frac{1}{3}\pi^{ab}tr\pi - \frac{\pi^{ab}tr\pi\psi^{12}}{6V^{2}}\biggr) \\ & + 
\frac{\sqrt{g}\psi}{V^{2/3}}\biggl(\grad^{a}\grad^{b}(N\psi^{3}) - 
g^{ab}\grad^{2}(N\psi^{3})\biggr) \\ & + 
\frac{N\psi^{3}\sqrt{g}}{V^{2/3}}\biggl(\grad^{a}\grad^{b}\psi + 
3g^{ab}\grad^{2}\psi\biggr)
\\ & 
+ 4\frac{\sqrt{g}}{V^{2/3}}g^{ab}\grad_{c}(N\psi^{3})\grad^{c}\psi - 
6\frac{\sqrt{g}}{V^{2/3}}\grad^{(a}(N\psi^{3})\grad^{b)}\psi 
\\ & + \grad_{c}\biggl(\pi^{ab}N^{c}\biggr) -  \pi^{bc}\grad_{c}N^{a} -  
\pi^{ac}\grad_{c}N^{b} \\ & - (\pi^{ab} - 
\frac{1}{2}g^{ab}tr\pi)\grad_{c}\xi^{c} - 
\frac{2}{3}\frac{\sqrt{g}\psi^{6}g^{ab}}{V^{2/3}}C \end{split} \eeq
where
\beq C = \biggl<N\sqrt{g}\psi^{4}\biggl(\rp + 
\frac{\psi^{4}(trp)^{2}}{4V^{2/3}}\biggr)\biggr> \eeq
and $\biggl<A\biggr>$ is the usual notion of global average given by
\beq \biggl<A\biggr> = 
\frac{\int \sqrt{g}A \,\, d^{3}x}{\int \sqrt{g} \,\, d^{3}x} \eeq
\\ \\
We can again take the time derivative of $trp$ and find yet again that
\beq \frac{\partial trp}{\partial t} = 0 \eeq
Thus, our dynamic data will once again be $\{g_{ab},\sigma^{ab}\}$ and so we 
want to find the evolution equation for $\sigma^{ab}$ again. Slogging through 
we get
\beq
\begin{split} \frac{\partial\sigma^{ab}}{\partial t} = & - 
\frac{N\sqrt{g}\psi^{4}}{V^{2/3}}\biggl(R^{ab} - 
\frac{1}{3}g^{ab}\biggl(\rp\biggr)\biggr) - \frac{2NV^{2/3}}{\sqrt{g}
\psi^{4}}\sigma^{ac}\sigma^{b}_{\;\;c} \\ & + \frac{\sqrt{g}\psi}{V^{2/3}}
\biggl(\grad^{a}\grad^{b}(N\psi^{3}) - 
\frac{1}{3}g^{ab}\grad^{2}(N\psi^{3})\biggr) \\ & + 
\frac{N\sqrt{g}\psi^{3}}{V^{2/3}}\biggl(\grad^{a}\grad^{b}\psi + 
\frac{7}{3}g^{ab}\grad^{2}\psi\biggr)
\\ & + \frac{4\sqrt{g}}{V^{2/3}}g^{ab}\grad_{c}(N\psi^{3})\grad^{c}\psi - 
\frac{6\sqrt{g}}{V^{2/3}}\grad^{(a}(N\psi^{3})\grad^{b)}\psi 
\\ & + \grad_{c}\biggl(\sigma^{ab}N^{c}\biggr) -  \sigma^{bc}\grad_{c}N^{a} - 
 \sigma^{ac}\grad_{c}N^{b} \\ & - \sigma^{ab}\grad_{c}\xi^{c} + 
\frac{N\psi^{8}}{3\sqrt{g}V^{4/3}}\sigma^{ab}tr\pi \end{split} \eeq
Note that the term with $C$ has dropped out.
\\ \\
The physical representation is achieved either by the naive substitution of 
$\psi = 1$ and $\grad_{c}\xi^{c} = 0$ or by doing it the longer more correct 
way. The result is the same in either case. The momentum is
\beq \pi^{ab} = - \frac{\sqrt{g}}{V^{2/3}}S^{ab} + 
\frac{2}{3}\sqrt{g}g^{ab}trKV^{4/3} \eeq
Thus
\beq \label{oops} \sigma^{ab} = -\frac{\sqrt{g}}{V^{2/3}}S^{ab} \;\; \text{and}
 \;\; tr\pi = 2\sqrt{g}(trK)V^{4/3} \eeq
The constraints are
\beq \sigma^{ab}\sigma_{ab} - \frac{1}{6}\frac{(tr\pi)^{2}}{V^{2}} = 
\frac{gR}{V^{4/3}} \eeq
\beq \grad_{b}\pi^{ab} = 0 \eeq
\beq \label{cc2} \grad_{c}trp = 0 \eeq
\beq NR -\grad^{2}N + \frac{N(trp)^{2}}{4V^{2/3}} =  C \eeq
where we now have 
$C = \biggl<N\biggl(R + \frac{(trp)^{2}}{4V^{2/3}}\biggr)\biggr>$. The 
evolution equations are
\beq \delg = \frac{2NV^{2/3}}{\sqrt{g}}\biggl(\sigma_{ab} - 
\frac{g_{ab}tr\pi}{6V^{2}}\biggr) + (KN)_{ab} \eeq
and
\beq
\begin{split} \frac{\partial\sigma^{ab}}{\partial t} = & - 
\frac{N\sqrt{g}}{V^{2/3}}\biggl(R^{ab} - 
\frac{1}{3}g^{ab}R\biggr) - \frac{2NV^{2/3}}{\sqrt{g}
}\sigma^{ac}\sigma^{b}_{\;\;c} \\ & + \frac{\sqrt{g}\psi}{V^{2/3}}
\biggl(\grad^{a}\grad^{b}N - \frac{1}{3}g^{ab}\grad^{2}N\biggr) \\ & + 
\grad_{c}\biggl(\sigma^{ab}N^{c}\biggr) - \sigma^{bc}\grad_{c}N^{a} - 
\sigma^{ac}\grad_{c}N^{b} \\ & + \frac{N}{3\sqrt{g}V^{4/3}}\sigma^{ab}tr\pi 
\end{split} \eeq
\section{The Volume}
This theory was inspired by the need to recover expansion. After all this work,
 have we succeeded? The time derivative of the volume is
\beq \begin{split} \frac{\partial V}{\partial t} = & 
\int\frac{1}{2}\sqrt{g}g^{ab}\delg\;\;d^{3}x \\ = & - 
\int\frac{1}{2}\frac{N\sqrt{g}trp}{V^{4/3}}\;\;d^{3}x \\ = & - 
\frac{trp\Big<N\Big>}{2V^{1/3}} \end{split} \eeq
Thus, we have recovered expansion. The big test of the compact 
without boundary theory presented here will be to study the 
cosmological solutions and this will be the focus of later work. 
Later in this paper we shall examine the consequences for the 
cosmological constant.
\section{Jacobi Action}
For completeness let's find the Jacobi action for the compact theory. Without 
going through each step let's simply require homogeneity in $\psi$. Recall that the Jacobi 
action for the non-compact theory was given by
\beqq S = \underline{+}\int d\lambda\int\sqrt{g}\psi^{4}\sqrt{\rp}\sqrt{T}d^{3}x
\eeqq
(\ref{nct}) where $ T = \biggl(\Sigma^{ab}\Sigma_{ab} - 
\frac{2}{3}\psi^{-12}(tr\beta)^{2}\biggr)$ where $\Sigma^{1b} = - 2NS^{ab}$ and
 $\beta^{ab} = - 2NB^{ab}$. Applying the homogeneity requirement gives
\beq S = \underline{+}\int d\lambda\int
\frac{\sqrt{g}\psi^{4}\sqrt{\rp}\sqrt{T}d^{3}x}{V^{2/3}} \eeq
where $ T = \biggl(\Sigma^{ab}\Sigma_{ab} - 
\frac{2}{3}\psi^{-12}(tr\beta)^{2}V^{2}\biggr) $.
\\ \\
Everything else emerges as before.
\section{Comparison with GR}
In the earlier ``static'' conformal theory we saw that the labelling
\beq \widehat{\pi^{ab}} = V^{2/3}\pi^{ab} \eeq
made the theory appear incredibly similar to GR. A similar labelling is 
possible here. Define
\beq \widehat{\sigma^{ab}} = V^{2/3}\sigma^{ab} \eeq
and
\beq \widehat{tr\pi} = \frac{tr\pi}{V^{1/3}} \eeq
With this rebelling the constraints are
\beq \widehat{\sigma^{ab}}\widehat{\sigma_{ab}} - 
\frac{1}{6}(\widehat{tr\pi})^{2} = gR \eeq
\beq \grad_{b}\widehat{\pi^{ab}} = 0 \eeq
\beq \grad_{c}\widehat{trp} = 0 \eeq
and the lapse-fixing equation is
\beq NR - \grad^{2}N + \frac{N(\widehat{trp})^{2}}{4} = C \eeq
where $C = \biggl<N\biggl(R + \frac{(\widehat{trp})^{2}}{4}\biggr)\biggr>$. 
These are \emph{identical} to GR in the CMC gauge. The evolution equations are
\beq \delg = \frac{2N}{\sqrt{g}}\biggl(\widehat{\sigma_{ab}} - 
\frac{g_{ab}\widehat{tr\pi}}{6V}\biggr) + (KN)_{ab} \eeq
and
\beq
\begin{split} \frac{\partial\widehat{\sigma^{ab}}}{\partial t} = & 
V^{2/3}\frac{\partial\sigma^{ab}}{\partial t} + 
\frac{2}{3V^{1/3}}\frac{\partial V}{\partial t} \\ = & - 
N\sqrt{g}\biggl(R^{ab} - \frac{1}{3}g^{ab}R\biggr) - 
\frac{2N}{\sqrt{g}}\widehat{\sigma^{ac}}\widehat{\sigma^{b}_{\;\;c}} \\ & + 
\sqrt{g}\psi\biggl(\grad^{a}\grad^{b}N - \frac{1}{3}g^{ab}\grad^{2}N\biggr) \\ 
& + \grad_{c}(\widehat{\sigma^{ab}}N^{c}) - 
\widehat{\sigma^{bc}}\grad_{c}N^{a} - 
\widehat{\sigma^{ac}}\grad_{c}N^{b} \\ & + 
\frac{\biggl(N - \Big<N\Big>\biggr)}{3\sqrt{g}V}
\widehat{\sigma^{ab}}\widehat{tr\pi} \end{split} \eeq
There are very few differences between these and those of GR (\ref{blugh}) and (\ref{yuch}).
\section{Constraint Propagation}
Of course, for consistency, we need the constraints to be preserved in time. It turns out that 
we have this here with one final restriction. The scalar curvature must be spatially constant. 
This condition is enough then for full constraint preservation.
\section{Time}
In his work on the initial value formulation of general relativity York \cite{yor} introduced 
the following time parameter (the York time)
\beq \tau = \frac{2}{3}trp \eeq
In this theory we have that $trp$ is identically constant. Thus it cannot be 
used as a notion of time. We note now though that unlike in GR, for us the 
volume is monotonically increasing. Of course, the volume is constant on any 
hypersurface by definition and so the volume provides a good notion of time in 
this theory. This may be extremely beneficial in a quantisation program.
\section{Light Cones}
So far the theory is quite promising. There are a number of things that must 
carry over from GR though if it is to be taken seriously. One of these is that 
the speed of propagation of the wave front must be unity (the speed of light). 
The easiest way to check this is to consider the evolution equations. Let's 
consider the case in GR briefly. The corresponding case in the conformal theory
 will work in almost exactly the same way.
\\ \\
The evolution equation for $g_{ab}$ in GR is
\beq \delg = \frac{2N}{\sqrt{g}}\biggl(\pi_{ab} - 
\frac{1}{2}g_{ab}tr\pi\biggr) + (KN)_{ab} \eeq
Inverting this we find that
\beq \pi_{ab} = 
\frac{\sqrt{g}}{2N}\frac{\partial g_{ab}}{\partial t} \eeq
We will be working here to leading order in the derivatives which is the reason
 for only omitting the other terms. Differentiating both sides gives
\beq \frac{\partial\pi_{ab}}{\partial t} = \frac{\sqrt{g}}{2N}
\frac{\partial^{2}g_{ab}}{\partial t^{2}} \eeq
Now substituting this into the evolution equation for $\pi_{ab}$ gives us
\beq \label{sol} \frac{\sqrt{g}}{2N}\frac{\partial^{2}g_{ab}}{\partial t^{2}} =
 - N\sqrt{g}\biggl(R_{ab} - \frac{1}{2}g_{ab}R\biggr) \eeq
(Note: The alternate form of the evolution equation is used here with the 
factor of $\frac{1}{2}$ on $R$.)
\\ \\
Now,
\beq \biggl(R_{ab} - \frac{1}{2}g_{ab}R\biggr) = \frac{1}{2}g^{cd}
\biggl[g_{bd,ac} + g_{ac,bd} - g_{ab,cd} - g_{cd,ab} - g_{ab}g^{ef}
\biggl(g_{ec,fd} - g_{ef,cd}\biggr)\biggr] \eeq
again only using leading order in the derivatives. We are concerned with the 
transverse traceless part of $g_{ab}$ which we'll label as $g_{ab}^{TT}$. The 
only relevant part is then
\beq - \frac{1}{2}g^{cd}g^{TT}_{ab,cd} \eeq
which we'll write as
\beq -\frac{1}{2}\frac{\partial^{2}g^{TT}_{ab}}{\partial x^{2}} \eeq
All the other terms are canceled either through the transverse or traceless 
properties. Using only the $TT$ part in the time derivatives also gives us
\beq \frac{1}{2N^{2}}\frac{\partial^{2}g^{TT}_{ab}}{\partial t^{2}} = 
\frac{1}{2}\frac{\partial^{2}g^{TT}_{ab}}{\partial x^{2}} \eeq
This is a wave equation with wave speed $1$. Thus we get gravitational 
radiation! Various details are omitted here but the essence of the idea is 
quite clear. Let's consider the conformal theory. We'll use the compact without
 boundary theory (that is, the one with the volume terms).
\\ \\
The evolution equation for $g_{ab}$ can be inverted to get
\beq \sigma_{ab}\frac{\sqrt{g}}{2NV^{2/3}}\delg + ... \eeq
Differentiating both sides gives
\beq \frac{\partial\sigma_{ab}}{\partial t} = \frac{\sqrt{g}}{2NV^{2/3}}
\frac{\partial^{2}g_{ab}}{\partial t^{2}} \eeq
again, working only to leading order in the derivatives. Substituting this into
 the evolution equation for $\sigma_{ab}$ gives us
\beq \frac{\sqrt{g}}{2NV^{2/3}}
\frac{\partial^{2}g_{ab}}{\partial t^{2}} = 
\frac{N\sqrt{g}}{V^{2/3}}\biggl(R_{ab} - \frac{1}{2}g_{ab}R\biggr) \eeq
The volume terms cancel and we are left with the same equation as (\ref{sol}) 
above. In exactly the same way this becomes
\beq \frac{1}{N^{2}}\frac{\partial^{2}g^{TT}_{ab}}{\partial t^{2}} = 
\frac{\partial^{2}g^{TT}_{ab}}{\partial x^{2}} \eeq
Yet again, we have found a wave equation with speed $1$. Thus we have recovered
 gravitational radiation with wavefronts propagating at the speed of light. All
 is still well.
\section{Matter and Cosmology}
The issues of coupling of matter and of cosmology will be treated in detail in 
forthcoming articles. Clearly, the theory here is incredibly restrictive 
cosmologically. We need a spatially constant scalar curvature an identically 
constant $trp$ and a monotonically changing volume. However, there is one 
interesting result which is easily found here regarding the cosmological 
constant.
\subsection{The Cosmological Constant Problem}
This is probably the best known problem of the so called standard cosmology. In
 GR we have the following. Taking the interpretation of the cosmological 
constant $\Lambda$ as a vacuum energy there is a discrepancy of at least 
$10^{120}$ orders of magnitude between the theoretically predicted value and 
the measured value today. That is
\beq \frac{\Lambda_{Pl}}{\Lambda_{0}} \geq 10^{120} \eeq
where the subscripts Pl and $0$ refer to Planck scales and today respectively. 
In GR the cosmological constant appears with the scalar curvature in the form 
$R + \Lambda$. However, in the 
new theory here it appears with a volume coefficient in the form $R + 
\frac{\Lambda}{V^{2/3}}$. Thus we are concerned with the ratio of 
$\frac{\Lambda}{V^{2/3}}$ at the Planck scale and today. We take the value of 
$\Lambda$ to be identically constant and so we wish to consider 
$\biggl(\frac{V_{0}}{V_{Pl}}\biggr)^{2/3}$. We take the radius of the universe 
at the Planck scale to be the Planck length which is approximately $10^{-35}$m.
 Today, the radius of the universe today is at least $10^{26}$m (the radius of 
the observed universe). Thus the ratio we are considering is
\beq \biggl(\frac{V_{0}}{V_{\text{Pl}}}\biggr)^{2/3} \geq 
\biggl(\frac{10^{26}}{10^{-35}}\biggr)^{2} = 10^{122} \eeq
There is no cosmological constant problem!
\\ \\
It should be pointed out that this is fundamentally different from postulating 
a ``time-varying cosmological constant''. The constant enters at the same level
 in his theory as in GR and it is the behaviour of the scalar curvature which 
changes things.
\\
As stated earlier, the full cosmological implications will be treated 
elsewhere. Such a restrictive theory is in general quite attractive providing 
definite testable predictions with relative ease. Indeed, looking at the above 
treatment of the cosmological constant in reverse as a prediction on the 
magnitude of $\Lambda$ is already one new prediction which seems to be 
satisfied experimentally!
\section{Comments}
\subsection{Derivations}
Naturally, as one would expect there exist alternative derivations of the 
theory. There are two in particular which are quite interesting. One of these 
will form part of a future paper. The other is quite straightforward and can be
 described quite easily.
\\
Consider the Hamiltonian of GR. It is
\beq \mathcal{H}_{GR} = \frac{N}{\sqrt{g}}\biggl(\pi^{ab}\pi_{ab} - 
\frac{1}{2}(tr\pi)^{2} - gR\biggr) -2N_{a}\grad{b}\pi^{ab} \eeq
We wish to construct a theory which is invariant under both diffeomorphisms and
 conformal transformations. Consider now the momentum constraint
\beq \grad_{b}\pi^{ab} = 0 \eeq
This implements diffeomorphism invariance. We need it to be conformally 
invariant also under the transformation
\beq g_{ab} \gt \omega^{4}g_{ab} \eeq
There are only two possibilities which satisfy this.
\\
(i) $\sigma^{ab} \gt \omega^{-4}\sigma^{ab}$ and $tr\pi = 0$; \nl
(ii) $\sigma^{ab} \gt \omega^{-4}\sigma^{ab}$, $\grad_{c}tr\pi = 0$ and $trp = \frac{tr\pi}{\sqrt{g}} \gt trp$.
\\
The first case here leads to the original conformal theory. The second is 
exactly the conformal transformation required to reproduce the new theory. 
Performing this transformation on the Hamiltonian constraint of GR (with 
conformal factor $\psi$) leads to the desired Hamiltonian constraint. Then 
finally, in a Dirac type procedure \cite{dir} we add the new constraint 
$\grad_{c}tr\pi = 0$ to the Hamiltonian with a Lagrange multiplier to obtain 
the full Hamiltonian of the theory. Thus, the theory is found in a simple and 
natural way.
\subsection{Quantisation}
The theory has several attractive features from a quantisation point of view. 
The configuration space is no longer simply superspace but has been reduced to 
conformal superspace plus a constant ($trp$). There is a physically preferred 
slicing and the volume of the universe emerges as a good notion of time. These 
point to possible benefits of a quantisation program for the theory. In 
particular, the theory may shed light on the problem of time in quantum 
gravity. Thus, regardless of its fate as a competitor to GR this theory may 
teach some valuable lessons.
\subsection{Recent Developments}
Since this was written further work by the author and collaborators has led to a first principles derivation of the full York method of general relativity \cite{abfko}. This is accomplished by only allowing invariance with respect to volume preserving conformal tranformations rather than general conformal transformations as in this work.
\section{Acknowledgements}
I wish to thank Edward Anderson, Julian Barbour, Brendan Foster and Niall \'{O} Murchadha for 
many valuable discussions. I also wish to thank the anonymous adjudicator for helpful comments 
and careful reading. This work was partially supported by Enterprise Ireland.


\vspace{.35cm}


\vspace{.2cm}

\begin{thebibliography}{160mm}
\bibitem[1]{abfom} E. Anderson, J. Barbour, B.Z. Foster, N. \'{O} Murchadha, 
\emph{Class. Quantum Grav.} $\mathbf{20}$ 1571 (2003), preprint gr-qc/0211022.
\bibitem[2]{bk} B. Kelleher, \emph{Class. Quantum Grav.} 
$\mathbf{21}$ 483 (2004), preprint gr-qc/0307091.
\bibitem[3] {yor}  J.W. York, \emph{Phys. Rev. Lett.} $\mathbf{26}$, 1656 
(1971); J.W. York, \emph{Phys. Rev. Lett.} $\mathbf{28}$ 1082 (1972); 
J. W. York, \emph{J. Math. Phys.} $\mathbf{14}$ 456 (1973). 
\bibitem[4] {bsw}  R.F. Baierlein, D. Sharp and J.A. Wheeler, 
\emph{Phys. Rev. Lett.} $\mathbf{126}$, 1864 (1962).
\bibitem[5] {adm}  R. Arnowitt, S. Deser, and C. Misner, 
\emph{Gravitation: an Introduction to Current Research}, ed. L. 
Witten, (Wiley, New York, 1962).
\bibitem[6]{dir} P.A.M. Dirac, \emph{Lectures on Quantum Mechanics} 
(Yeshiva University, NY, 1964). 
\bibitem[7] {abfko} E. Anderson, J. Barbour, B.Z. Foster, B. Kelleher, 
N. \'{O} Murchadha, preprint gr-qc/0404099.
\end{thebibliography}
\end{document}